\documentclass[aps, prd, twocolumn, preprintnumbers, superscriptaddress, nofootinbib, floatfix]{revtex4-1}

\usepackage{amsmath}
\usepackage{bm}
\usepackage{amsfonts}
\usepackage{graphicx}
\usepackage{epstopdf}
\usepackage{hyperref}
\usepackage{array}
\usepackage[margin=1in,footskip=0.2in]{geometry}
\usepackage{color}

\enlargethispage{\baselineskip}

\hypersetup{backref,
colorlinks=true,
linkcolor=blue,
linktoc=page,
citecolor=blue,
urlcolor=blue}

\everymath{\displaystyle} 

\newcommand{\MP}{M_{\rm Pl}}

\newcommand{\PkO}{A_s}
\newcommand{\Pk}{\Delta^2_{\mathcal{R}}(k)}

\newcommand{\be}{\begin{equation}}
\newcommand{\ee}{\end{equation}}
\newcommand{\bea}{\begin{eqnarray}}
\newcommand{\eea}{\end{eqnarray}}

\begin{document}

\title{\bf The zoo plot meets the swampland: mutual (in)consistency of single-field inflation, string conjectures, and cosmological data}

\newcommand{\FIRSTAFF}{\affiliation{The Oskar Klein Centre for Cosmoparticle Physics, Department of Physics, Stockholm University, Roslagstullsbacken 21A, SE-10691 Stockholm, Sweden}}
\newcommand{\SECONDAFF}{\affiliation{The Nordic Institute for Theoretical Physics (NORDITA), Roslagstullsbacken 23, SE-10691 Stockholm, Sweden}}
\newcommand{\THIRDAFF}{\affiliation{Department of Physics and Astronomy, Uppsala University, L\"agerhyddsv\"agen 1, SE-75120 Uppsala, Sweden}}
\newcommand{\FOURTHAFF}{\affiliation{Department of Physics, University at Buffalo, Buffalo, NY 14260, USA}}

\author{William H. Kinney}
\email[Electronic address: ]{whkinney@buffalo.edu}
\FIRSTAFF
\FOURTHAFF
\author{Sunny Vagnozzi}
\email[Electronic address: ]{sunny.vagnozzi@fysik.su.se}
\FIRSTAFF
\SECONDAFF
\author{Luca Visinelli}
\email[Electronic address: ]{luca.visinelli@physics.uu.se}
\FIRSTAFF
\SECONDAFF
\THIRDAFF

\date{\today}
\preprint{NORDITA-2018-075}
\begin{abstract}
We consider single-field inflation in light of string-motivated ``swampland'' conjectures suggesting that effective scalar field theories with a consistent UV completion must have field excursion $\Delta \phi \lesssim \MP$, in combination with a sufficiently steep potential, $\MP V_\phi/V \gtrsim {\cal O}(1)$. Here, we show that the swampland conjectures are inconsistent with existing observational constraints on single-field inflation. Focusing on the observationally favoured class of concave potentials, we map the allowed swampland region onto the $n_S$-$r$ ``zoo plot" of inflationary models, and find that consistency with the Planck satellite and BICEP2/Keck Array requires $\MP V_\phi/V \lesssim 0.1$ and $-0.02 \lesssim \MP^2 V_{\phi\phi}/V < 0$, in strong tension with swampland conjectures. Extension to non-canonical models such as DBI Inflation does not significantly weaken the bound.
\end{abstract}

\maketitle

\section{The swampland-de Sitter conjectures}

Inflation is a postulated period of quasi-de Sitter expansion in the very early Universe (see e.g. Refs.~\cite{Guth:1979bh,Kazanas:1980tx,Starobinsky:1980te,Guth:1980zm,Linde:1981mu,Sato:1981ds,Mukhanov:1981xt,Izawa:1982cu,Albrecht:1982wi,Linde:1983gd} for early seminal work). Inflation provides an explanation for various problems arising in the standard Big Bang cosmology such as the observed homogeneity, flatness, and lack of relic monopoles. In addition, inflation provides a mechanism for generating the density perturbations whose existence we infer from the observation of temperature anisotropies in the Cosmic Microwave Background Radiation (CMBR)~\cite{Guth:1982ec,Hawking:1982cz, Starobinsky:1982ee,Bardeen:1983qw,Steinhardt:1984jj}: these density perturbations, generated when quantum fluctuations leave the Hubble radius during inflation, later grow under gravitational instability to form the observed cosmic large-scale structure. Among the simplest class of inflationary models are single-field models, wherein a dynamical scalar field (the ``inflaton") evolves under the influence of a nearly-flat potential, leading to an approximately constant expansion rate.

It would naturally be desirable to embed inflationary models within well-motivated high-energy UV-complete theories, with the latter possibly including a consistent description of quantum gravity: the behaviour of the inflaton field would then be captured by the effective field theory (EFT) given by the low-energy limit of this UV-complete theory. String theory naturally emerges as a potential candidate for such a UV-complete theory. While the huge landscape of string vacua is believed to lead to consistent EFTs, these are conjectured to be surrounded by a ``swampland" of  semi-classical EFTs which do not allow for a consistent theory of quantum gravity~\cite{Vafa:2005ui}.

Attempts to establish the conditions under which a given EFT does not lie in the swampland have led to a set of conjectures, such as the the weak-gravity conjecture~\cite{ArkaniHamed:2006dz} and, more recently, a set of additional swampland conjectures~\cite{Ooguri:2006in,Klaewer:2016kiy,Ooguri:2016pdq,Freivogel:2016qwc,Brennan:2017rbf,Obied:2018sgi}. In particular, it has long been noted that, while it is easy to obtain Minkowski and Anti-de Sitter vacua in string theory, the same cannot be said about de Sitter (dS) vacua, stable versions of which are notoriously extremely hard to obtain~\cite{Bena:2014jaa,Kutasov:2015eba,Moritz:2017xto,Danielsson:2018ztv}. This has raised the suspicion that theories featuring dS vacua reside in the swampland, rather than in the landscape~\cite{Dvali:2013eja,Obied:2018sgi}. This observation would obviously have profound implications for inflationary theories.

Here, we focus on two swampland conjectures~\cite{Obied:2018sgi} whose cosmological implications were recently studied in Ref.~\cite{Agrawal:2018own}. These two conjectures, which we shall refer to as ${\cal SC}_1$ and ${\cal SC}_2$ respectively, place constraints on the proper range traversed by scalar fields in field space, $\Delta \phi$, as well as on the logarithmic gradient of the scalar field potential $V(\phi)$,
\begin{eqnarray}
\frac{\vert \Delta \phi \vert}{M_{\rm Pl}} \lesssim \Delta \sim {\cal O}(1)\, \quad \quad \left ( {\cal SC}_1 \right ) \,, \label{sc1}\\
M_{\rm Pl}\frac{\vert V_{\phi} \vert}{V} \gtrsim c \sim {\cal O}(1)\, \quad \quad \left ( {\cal SC}_2 \right ) \,,
\label{sc2}
\end{eqnarray}
where $M_{\rm Pl} \simeq 2.4 \times 10^{18}\,{\rm GeV}$ is the reduced Planck mass, $V_{\phi}$ stands for $dV(\phi)/d\phi$, and $c$ is a positive constant of order unity whose actual value depends on the details of the compactification, but which in many string realizations is typically larger than $\sqrt{2}$ and never smaller than unity~\cite{Obied:2018sgi}. It is not hard to see how ${\cal SC}_2$ represents a hazard for inflationary theories, which typically require $V_{\phi} \ll V$ in order to sustain a sufficiently long period (of at least about 60 e-folds) of quasi-de Sitter expansion.

The two swampland conjectures ${\cal SC}_1$ and ${\cal SC}_2$ have received significant attention in the recent literature, and various follow-up works have examined their implications for cosmology (including but not limited to inflation and dark energy) and more generally fundamental physics. For an incomplete list of relevant papers, see e.g. Refs.~\cite{Agrawal:2018own,Andriot:2018wzk,Dvali:2018fqu,Banerjee:2018qey,Achucarro:2018vey,Garg:2018reu,Lehners:2018vgi,Kehagias:2018uem,
Dias:2018ngv,Denef:2018etk,Colgain:2018wgk,Roupec:2018mbn,Andriot:2018ept,Matsui:2018bsy,Ben-Dayan:2018mhe,Chiang:2018jdg,Heisenberg:2018yae,Damian:2018tlf,Gu:2018akj,Conlon:2018eyr,
Visinelli:2018utg}. Several of these works, in particular, noticed that single-field slow-roll inflation is in tension with criterion ${\cal SC}_2$~\cite{Agrawal:2018own,Achucarro:2018vey,Garg:2018reu,Kehagias:2018uem,Dias:2018ngv}, since the latter places a lower bound on the amplitude of primordial gravitational waves produced during inflation, parametrized by the tensor-to-scalar ratio $r$. It can be easily shown that ${\cal SC}_2$ in Eq.~(\ref{sc2}) implies $r>8c^2$ for single-field slow-roll inflation~\cite{Dias:2018ngv}, and since current constraints from the \textit{Planck} satellite in combination with the BICEP2/Keck Array ground-based small aperture telescopes suggest $r<0.07$ at 95\% confidence level (C.L.)~\cite{Array:2015xqh}, a value of the parameter $c \sim {\cal O}(1)$ would strongly violate such bound.

In this note, it is our goal to revisit the status of single-field slow-roll inflation in light of current observational data and the conjectured swampland criteria. However, unlike previous work which especially focused on the tensor-to-scalar ratio $r$, here we also consider the restrictions imposed by the second swampland criterion on the scalar spectral index $n_S$, including the second derivative of the potential. We consider for definiteness models where the potential of the inflaton is concave ($V_{\phi\phi} \equiv d^2V/d\phi^2<0$) rather than convex, since the former are observationally preferred over the latter.  This allows us to reach a conclusion on the inconsistency between single-field slow-roll inflation, the swampland conjectures, and observational data, which is stronger than those previously reached focusing only on $r$. In particular, we find that single-field slow-roll inflaton potentials with $c>0.1$ are ruled out at $>95\%$ C.L. under generic assumptions based on string theory. We also consider constraints on the curvature of the potential in light of the more recently proposed ``refined'' swampland criterion \cite{Ooguri:2018wrx}, and place a lower bound $-0.02 \lesssim \MP^2 V_{\phi\phi}/V$, in strong tension with the refined conjecture. We then briefly consider convex potentials, restricting ourselves to the case where the inflaton is of the Dirac-Born-Infeld (DBI) form in order to circumvent tight constraints from overproduction of tensor modes, finding that our conclusions are qualitatively unchanged.

The rest of this note is organized as follows. In Sec.~\ref{sec:si}, we briefly review the equations of motion in single-field slow-roll inflation and the definition of the slow-roll parameters. In Sec.~\ref{sec:observables}, we make the connection to the observables, namely the scalar spectral index $n_S$ and the tensor-to-scalar ratio $r$, focusing on concave potentials. In Sec.~\ref{sec:dbi}, we briefly consider the case of convex potentials where the inflaton is of the DBI form, and verify that our conclusions are qualitatively unchanged. Finally, in Sec.~\ref{sec:conclusions} we provide concluding remarks. Our main results are showcased in Fig.~\ref{fig:nsrplane}, where we make connection to the well-known ``zoo plot'' of inflationary models on the $n_S$-$r$ plane~\cite{Dodelson:1997hr}. 

\section{Single-field slow-roll inflation}
\label{sec:si}

We consider a generic single-field slow-roll inflationary model wherein the role of inflaton is played by a scalar field $\phi = \phi(x)$ moving under the influence of a potential $V = V(\phi)$. Within a Friedmann-Robertson-Walker metric, the equation of motion for the inflaton field is  expressed as
\begin{eqnarray}
	\ddot{\phi} + 3H\dot{\phi} + V_\phi = 0,
	\label{eq_motion}
\end{eqnarray}
where a dot indicates derivation with respect to cosmic time. On the other hand, the Friedmann equation for the expansion rate of the Universe is
\begin{eqnarray}
	H^2 = \frac{\rho}{3\MP^2} = \frac{1}{3\MP^2} \left ( \frac{1}{2}\dot{\phi}^2+V \right ),
	\label{friedmann}
\end{eqnarray}
where as previously $M_{\rm Pl}$ denotes the reduced Planck mass, while the term within parentheses on the right-hand side describes the total energy density of the system.

If the potential $V(\phi)$ is sufficiently flat and dominates over the kinetic energy of the inflaton, i.e. $\dot{\phi}^2 \ll V(\phi)$, inflation occurs, with the Hubble expansion rate being approximately constant. During this slow-roll regime, the higher time derivatives in Eq.~(\ref{eq_motion}) can be neglected, i.e. $\ddot{\phi} \ll H \,\dot{\phi}$. Therefore, Eqs.~(\ref{eq_motion},\ref{friedmann}) reduce to
\begin{eqnarray}
	\dot{\phi} &\simeq& -\frac{V_\phi}{3H}\,, \label{eq_motion_Slow_roll}\\
	H^2 &\simeq& \frac{V}{3\MP^2}, \label{friedmann_Sl}
\end{eqnarray}
where we hereafter use the symbol ``$\simeq$" to imply an approximate equality that holds during the slow-roll regime. From Eq.~\eqref{friedmann_Sl} it follows that a shallow potential $V$ gives rise to a nearly constant expansion rate $H$. The physics during slow-roll can be parametrized by a set of slow-roll parameters $\epsilon$, $\eta$ defined by~\cite{Liddle:1994dx}
\begin{eqnarray}
	\epsilon \equiv \frac{\MP^2}{2}\,\left(\frac{V_\phi}{V}\right)^2, \quad \eta \equiv \MP^2\,\frac{V_{\phi\phi}}{V}.
	\label{slow_roll_parameters}
\end{eqnarray}
Slow-roll is achieved as long as the conditions $\epsilon\,, \vert \eta \vert \ll 1$ are satisfied, whereas inflation ends when $\epsilon\,, \vert \eta \vert \sim {\cal O}(1)$.

\section{Connection to observables}
\label{sec:observables}

Within the inflationary scenario, scalar and tensor fluctuations are redshifted to superhorizon scales and later evolve into primordial perturbations in the density field as well as primordial gravitational waves, leaving an imprint in the CMBR anisotropy and on the large-scale structure~\cite{Mukhanov:1981xt,Guth:1982ec,Hawking:1982cz,Starobinsky:1982ee,Bardeen:1983qw,Steinhardt:1984jj}. The spectrum of the adiabatic density (scalar) perturbations generated during inflation is expressed by the dimensionless power spectrum $\Delta_{\mathcal{R}}^2(k)$, which describes the contribution to the total variance of primordial curvature perturbations at a given scale per logarithmic interval in $k$ and is given as~\cite{Kosowsky:1995aa,Leach:2002dw,Liddle:2003as}
\begin{eqnarray}
	\Pk \equiv \frac{k^3\,P_{\mathcal{R}}(k)}{2\pi^2} = \PkO\,\left(\frac{k}{k_0}\right)^{n_S-1}\,,
	\label{curvature_perturbations}
\end{eqnarray}
where $\PkO \sim 2.2\times 10^{-9}$ is the amplitude of the primordial scalar power spectrum. The scalar spectral index $n_S$ parametrizes the mild dependence of the power spectrum on the co-moving wavenumber $k$, with $n_S = 1$ corresponding to a scale-invariant power spectrum. In single-field slow-roll inflation,  the scalar spectral index is given by (see e.g. Ref.~\cite{Baumann:2009ds})
\begin{eqnarray}
	n_S - 1 \approx -6\epsilon + 2\eta\,,
	\label{eq:scalartilt}
\end{eqnarray}
to leading order in the slow-roll parameters.

\begin{figure}[!h]
\begin{center}
\includegraphics[width=\linewidth]{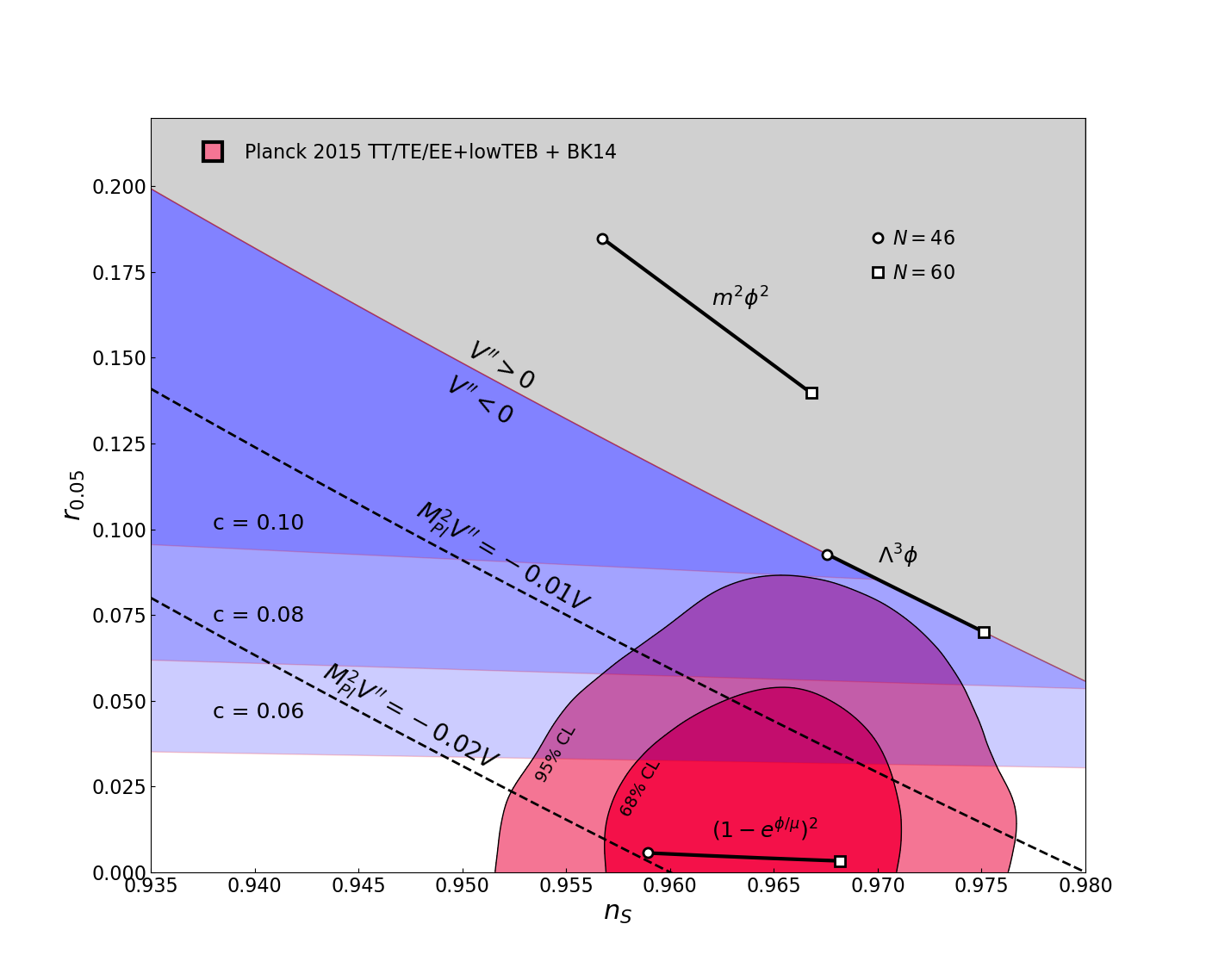}
\caption{Regions in the $n_S$-$r$ plane allowed by data (red) and the swampland conjectures (blue) for various values of the parameter $c$ and with $c' = 0$. In particular, the red contours denote two-dimensional probability contours at 68\% (dark red) and 95\%~C.L. (light red) in light of measurements of temperature and polarization anisotropies from \textit{Planck} and measurements of degree-scale B-mode polarization from BICEP2/Keck Array. The region in the blue shaded contours is the region of parameter space allowed for single-field slow-roll inflation by the swampland conjectures, for various values of the parameter $c$ appearing in Eq.~(\ref{sc2}). For definiteness we have focused on concave inflationary potentials ($V_{\phi\phi}<0$), whose parameter space lies below the gray shaded region. It is clear that $c \lesssim {\cal O}(0.1)$ is required for the predictions of single-field slow-roll inflation to be consistent with data, violating the string-based expectation that $c$ should be of order unity (the pivot scale plotted is $k_* = 0.05\, h\,\mathrm{Mpc}^{-1}$, and the calculated quantities $n_S$ and $r$ have been rescaled from $k_* = 0.002\, h\,\mathrm{Mpc}^{-1}$). The dashed lines show contours for different values of the second slow roll parameter $\eta = \MP^2 V_{\phi\phi}/V$: the region below the dashed lines is constrained by the refined swampland conjecture, Eq.~\eqref{sc2r}, with $c' = 0.01$ and $c' = 0.02$, respectively.}
\label{fig:nsrplane}
\end{center}
\end{figure}

Recalling the definition of $\epsilon$ and $\eta$ in Eq.~(\ref{slow_roll_parameters}), we are now ready to examine the implications of the second swampland criterion ${\cal SC}_2$ [Eq.~(\ref{sc2})] for the allowed values of the scalar spectral index $n_S$. It is clear that Eq.~(\ref{sc2}) in combination with the definition in Eq.~(\ref{slow_roll_parameters}) directly constrains the first slow-roll parameter $\epsilon$. Regarding the second derivative of the potential, quantum corrections typically force $\MP^2\,\left\vert V_{\phi\phi}\right\vert \sim V$ in the absence of a symmetry, which is the well-known ``$\eta$ problem''~\cite{Copeland:1994vg}. This statement has been recently formalised and included in a refined version, here ${\cal SC}_{2r}$, of the second swampland condition in Eq.~\eqref{sc2} (see Ref.~\cite{Ooguri:2018wrx}). For a single-field potential, ${\cal SC}_{2r}$ reads 
\be
	M_{\rm Pl}\frac{\vert V_{\phi} \vert}{V} \gtrsim c \sim {\cal O}(1)\, \quad\hbox{or} \quad M_{\rm Pl}^2\frac{ V_{\phi\phi} }{V} \lesssim -c'\, ,
\label{sc2r}
\ee
where $c'$ is another constant of order unity. Using the condition ${\cal SC}_{2r}$, we then immediately obtain
\begin{eqnarray}
	\epsilon \gtrsim \frac{c^2}{2}\,\quad \hbox{or}
		\label{epsiloninequality}
	\quad \eta \lesssim - c'\,.
		\label{etainequality}
\end{eqnarray}
The first of the two inequalities in Eqs.~(\ref{epsiloninequality}) has been already shown in a number of recent papers~\cite{Kehagias:2018uem, Dias:2018ngv, Matsui:2018bsy, Ben-Dayan:2018mhe}, whereas the second of Eqs.~(\ref{etainequality}) is motivated by both the $\eta$ problem and by the refined version of the swampland conjecture ${\cal SC}_{2r}$. Observational constraints demand $\eta \leq 0$, corresponding to concave potentials for single-field slow-roll inflation, $V_{\phi\phi} < 0$ (see Fig. \ref{fig:nsrplane}, since convex potentials lead to an overproduction of tensor modes for observationally allowed values of $n_S$). The bounds in Eqs.~(\ref{epsiloninequality}) then have definite implications for the scalar spectral index. Combining with Eq.~(\ref{eq:scalartilt}), we obtain the bound
\be
	1-n_S \gtrsim [3c^2; 2c']\,,
	\label{nsinequality}
\ee
where the term in square brackets is $3c^2$ for the case ${\cal SC}_{2}$, or $2c'$ for the case ${\cal SC}_{2r}$. In addition, single-field slow-roll inflation predicts the consistency relation $r = 16\epsilon$ which, combined with the constraint in Eqs.~(\ref{epsiloninequality}), leads to~\cite{Kehagias:2018uem, Dias:2018ngv, Matsui:2018bsy, Ben-Dayan:2018mhe}
\be
	r > 8c^2\,.
	\label{rinequality}
\ee
Measurements of the CMBR temperature and polarization anisotropies from the \textit{Planck} satellite instead set $n_S=0.9645\pm0.0049$ (\textit{Planck} TT,TE,EE+lowP dataset combination), with small shifts of ${\cal O}(0.001)$ when external data (such as Baryon Acoustic Oscillation distance measurements) are also included, or when different assumptions are made concerning the mass spectrum of massive neutrinos~\cite{Gerbino:2016sgw}. Regardless of the prior used, it is safe to assert that cosmological observations generically require $1-n_S \simeq 0.04$ and $r<0.1$ measured at the quadrupole with $k_* = 0.002\ h\,{\rm Mpc}^{-1}$. The bounds in Eqs.~\eqref{nsinequality} and~\eqref{rinequality} imply $c \lesssim {\cal O}(0.1)$ when compared with the observational results, which is in clear contrast with the string-based expectation that $c$ should be of order unity if not greater.

To make contact with observations, we overlay the constraints imposed by Eqs.~\eqref{nsinequality} and~\eqref{rinequality} on the usual ``zoo plot'', the $n_S$-$r$ plane in Fig.~\ref{fig:nsrplane}, for selected values of the parameter $c=0.06\,,0.08\,,0.10$ and for $c' = 0$. The region selected by the swampland constraints is shaded in blue, whereas the red contours correspond to the 68\% and 95\%~C.L. allowed regions when constrained with the \textit{Planck} 2015 TT,TE,EE+lowTEB temperature and polarization data~\cite{Adam:2015rua,Ade:2015xua}, and the BICEP2/Keck Array 2014 combined polarization data~\cite{Ade:2015fwj}. The allowed contours are calculated numerically using a Markov Chain Monte Carlo method with the \texttt{CosmoMC} sampler~\cite{Lewis:2002ah}, which makes use of the \texttt{CAMB} Boltzmann code. We fit to a seven-parameter $\Lambda$CDM+$r$ model with the following parameters:~\footnote{The data sets themselves contain multiple internal parameters, which we do not list here.} 
\begin{itemize}
\item{Baryon density $\Omega_{\rm b} h^2$.}
\item{Dark matter density $\Omega_{\rm C} h^2$.}
\item{Angular scale of acoustic horizon $\theta$ at decoupling.}
\item{Reionization optical depth $\tau$.}
\item{Power spectrum normalization $A_s$.}
\item{Tensor-to-scalar ratio $r$, calculated at a pivot scale of $k = 0.05\ h \mathrm{Mpc}^{-1}$.}
\item{Scalar spectral index $n_{\rm S}$.}
\end{itemize}
Curvature $\Omega_{\rm k}$ is set to zero, and the Dark Energy equation of state is fixed at $w = -1$. We fix the number of neutrino species at $N_\nu = 3.046$, with one massive neutrino with mass $m_\nu = 0.06\ {\rm eV}$. For these constraints, we run 8 parallel chains with Metropolis-Hastings sampling, and use a convergence criterion of $R - 1 < 0.05$ for the Gelman-Rubin parameter $R$. It is clear from the figure that $c=0.1$ is already in tension with the data, whereas $c$ of order unity (not shown) is excluded at extremely high significance. Constrains from the refined swampland conjecture are also in significant tension with the data, see the dashed lines in Fig.~\ref{fig:nsrplane} with $c' = 0.01$ and $c' = 0.02$, respectively.

We therefore obtain limits from data on conjectures \eqref{sc2} and \eqref{sc2r}, respectively of
\begin{eqnarray}
\MP \frac{V_\phi}{V} &\lesssim& 0.1\cr
\MP^2 \frac{V_{\phi\phi}}{V} &\lesssim& 0.02.
\end{eqnarray}
There is no significant constraint arising from the constraint \eqref{sc1} on the field excursion, since the Planck+BICEP2/Keck Array upper bound of $r < 0.07$ corresponds to a bound on the field excursion \cite{Lyth:1996im}.
\begin{equation}
\Delta\phi \sim \MP \sqrt{r} < 0.26,
\end{equation}
which is consistent with swampland conjectures. 

\section{Convex Potentials and DBI Inflation}
\label{sec:dbi}

We have so far confined ourselves to a discussion of concave potentials, for which $V_{\phi\phi} < 0$. What about convex potentials, with $V_{\phi\phi} > 0$? Such models can satisfy constraints on the scalar spectral index $n_S$ even with $\eta \sim \epsilon$, since for $\eta > 0$, the quantity $n_S - 1$ in Eq.~\eqref{eq:scalartilt} might remain small via a cancellation between the opposite-sign $\epsilon$ and $\eta$ terms. However, in the case of canonical single-field slow-roll inflation, potentials with $V_{\phi\phi} > 0$ overproduce tensor perturbations, in conflict with data (Fig.~\ref{fig:nsrplane}). For example, the well-studied case of power-law inflation, with $V\left(\phi\right) \propto \exp({\phi/\MP})$, satisfies
\begin{eqnarray}
\MP \frac{V_\phi}{V} = \sqrt{2 \epsilon} = \mathrm{const.}
\end{eqnarray}
The scalar spectral index and the tensor-to-scalar ratio are respectively
\begin{eqnarray}
	n_S &=& 1 - 2 \epsilon,\\
	r &=& 16 \epsilon,
\end{eqnarray}
so that $n_S - 1 \sim 0.04$ requires $\epsilon \sim 0.02$, resulting in a tensor-to-scalar ratio of $r \sim 0.3$, which is excluded by Planck constraints to high significance, and is also inconsistent with the swampland conjectures, since $c \sim \sqrt{\epsilon} \sim 0.1$, similar to our constraint on concave potentials.   

However, these bounds are relaxed if the inflaton field is of the Dirac-Born-Infeld (DBI) form~\cite{Silverstein:2003hf}, in which an effective speed limit keeps the inflaton field near the top of $V(\phi)$ even if the potential is steep. In string theory models of inflation, the inflaton is a modulus parameter of a D-brane propagating in an AdS warped throat region of an approximate Calabi-Yau flux compactification. The propagation of fluctuations depend on the sound speed parameter $c_S$ which, in DBI inflation, is smaller than unity. This reduced sound speed has the effect of suppressing tensor production, with the tensor-to-scalar ratio given by
\begin{eqnarray}
	r \simeq 16 c_S \epsilon.
\end{eqnarray}
This means that non-canonical extensions of models with convex potentials can be brought into agreement with data in the case of $c_S \ll 1$ \cite{Tzirakis:2008qy,Bessada:2009pe}. In the case of $c_S < 1$, additional constraints come into play, in particular limits from Planck on primordial non-Gaussianity. DBI inflation predicts an amplitude in the equilateral mode for non-Gaussianity of
\begin{eqnarray}
	f_{\rm NL}^{\mathrm{Equil.}} = \frac{35}{108} \left(\frac{1}{c_S^2} - 1\right),
\end{eqnarray}
which is constrained by Planck to $f_{\rm NL}^{\mathrm{Equil.}} = -4 \pm 43$ to $1 \sigma$. Then the $2 \sigma$ upper bound of $f_{\rm NL}^{\mathrm{Equil.}} < 82$ gives a $2 \sigma$ {\it lower} bound on $c_S$ of \cite{Stein:2016jja}
\begin{eqnarray}
c_S > 0.0627. 
\end{eqnarray}
This can be combined with the upper bound from Planck and BICEP2/Keck Array on the tensor-to-scalar ratio of
\begin{eqnarray}
r = 16 c_S \epsilon < 0.07,
\end{eqnarray}
to give an upper bound
\begin{eqnarray}
\epsilon < 0.07 \rightarrow c < \sqrt{2 \epsilon} < 0.37.
\end{eqnarray}
Therefore, the combination of Planck upper bounds on both the tensor-to-scalar ratio and the amplitude of equilateral non-Gaussianity place limits on non-canonical models with $V_{\phi\phi} > 0$ slightly weaker than the constraints we obtained for the $V_{\phi\phi} < 0$ case from limits on $r$ and $n_S$, but still in strong tension with the swampland conjectures.

\section{Conclusions}
\label{sec:conclusions}

In this note, we have revisited the status of single-field inflation in light of the swampland conjectures Eqs.~(\ref{sc1},\ref{sc2}). Unlike previous related works~\cite{Agrawal:2018own,Achucarro:2018vey,Garg:2018reu,Kehagias:2018uem,Dias:2018ngv}, we have focused not only on the implications of the swampland conjectures for the allowed values of the tensor-to-scalar ratio $r$, but also for the scalar spectral index $n_S$, under the generic expectation that the swampland conjectures lead to $\epsilon \sim \mathcal{O}(c^2)$ and data favour $\eta < 0$. Our main result is shown in Fig.~\ref{fig:nsrplane}, where we showcase the regions in the $n_S$-$r$ plane allowed by the swampland conjectures for various values of the parameters $c$ and $c'$, and compare these to the regions allowed by current observational data. We clearly see that $c \lesssim {\cal O}(0.1)$ and $c' \lesssim {\cal O}(0.01)$ are required to obtain consistency between the predictions of single-field slow-roll inflation and data, at the cost of violating the string-based expectation that $c, c' \sim {\cal O}(1)$. Extension to DBI models results in a constraint of $c \lesssim 0.37$, still in significant tension with the swampland conjectures.

\begin{acknowledgments}
The authors acknowledge support by the Vetenskapsr\r{a}det (Swedish Research Council) through contract No. 638-2013-8993 and the Oskar Klein Centre for Cosmoparticle Physics. WHK is supported by the U.S. National Science Foundation under grant NSF-PHY-1719690. This work was performed in part at the University at Buffalo Center for Computational Research.
\end{acknowledgments}

\bibliography{Swampland_cqg_R1}

\end{document}